\documentclass[twocolumn,showpacs,preprintnumbers,prb]{revtex4-1}
\usepackage{eurosym,hyperref}
\usepackage{graphicx,bm,amsmath,amssymb}

\hypersetup{colorlinks=true,citecolor=blue,linkcolor=magenta}
\allowdisplaybreaks

\def\gz{\ifmmode{Z\hskip -4.8pt Z}
    \else{\hbox{$Z\hskip -4.8pt Z$}}\fi}

\newcommand{\be}{\begin{equation}}
\newcommand{\ee}{\end{equation}}
\newcommand{\bea}{\begin{eqnarray}}
\newcommand{\eea}{\end{eqnarray}}

\begin{document}

\title{Effective one-band models for the 1D cuprate 
Ba$_{2-x}$Sr$_x$CuO$_{3+\delta}$}

\author{A. E. Feiguin}
\affiliation{Physics Department, Northeastern University, Boston, MA 02115, USA}

\author{Christian Helman}
\affiliation{Centro At\'omico Bariloche and Instituto Balseiro, CNEA, GAIDI 8400 S.
C. de Bariloche, Argentina}

\author{A. A. Aligia}
\affiliation{Instituto de Nanociencia y Nanotecnolog\'{\i}a CNEA-CONICET, GAIDI,
Centro At\'{o}mico Bariloche and Instituto Balseiro, 8400 Bariloche, Argentina}

\begin{abstract}
We consider a multiband Hubbard model $H_m$ for Cu and O orbitals in
Ba$_{2-x}$Sr$_x$CuO$_{3+\delta}$ similar to the tree-band model for two-dimensional (2D) cuprates. The hopping parameters are 
obtained from maximally localized Wannier functions derived from  \textit{ab initio} calculations. 
Using the cell perturbation method, we derive both a generalized $t-J$ model $H_{tJ}$ and a one-band Hubbard model 
$H_{H}$ to describe the low-energy physics of the system. 
$H_{tJ}$ has the advantage of having a smaller relevant Hilbert space, facilitating numerical calculations, while additional terms should be included in $H_{H}$ to 
accurately describe the multi-band physics of $H_m$. Using $H_{tJ}$ and  DMRG, we
calculate the wave-vector resolved photoemission and discuss the relevant features in comparison with recent experiments. In agreement with previous calculations, we find that the addition of an attractive nearest-neighbor interaction of the order of the nearest-neighbor hopping shifts weight from the $3 k_F$ to the holon-folding branch. Kinetic effects also contribute to this process. 

\end{abstract}

%\pacs{71.27.+a, 74.72.-h}

%72.15.Qm Scattering mechanisms and Kondo effect
%75.20.Hr Local moment in compounds and alloys; Kondo effect, valence fluctuations, heavy fermions 
%71.27.+a Strongly correlated electron systems; heavy fermions
%74.72.-h Cuprate superconductors (high-Tc and insulating parent compounds)

\maketitle

\section{Introduction}

\label{intro}

After more than three decades from the discovery of high-$T_c$ superconductivity, the pairing mechanism is still not fully
understood, although it is believed that it is related
to spin fluctuations originating from the effective 
Cu-Cu superexchange $J$, and complicated by the existence
of several phases competing with superconductivity
\cite{keimer,fradkin,maho,dong}.
However, there is consensus that for energies below an energy scale of the order of 1 eV, the physics of the 
two-dimensional (2D) superconducting
cuprates is described by the three-band Hubbard model\cite{varma,eme} $H_{3b}$, which contains the 3d$_{x^{2}-y^{2}}$ orbitals of Cu and the 2p$_{\sigma }$ orbitals of O \footnote{To explain some Raman and photoemission experiments at higher
energies, other orbitals should be included (see for example Ref. \onlinecite{raman}), but we can neglect them in this work)}. 

More recently, one-dimensional (1D) cuprates have attracted
a great deal of attention, in particular because numerical techniques in 1D are 
more powerful and also field-theoretical methods like bosonization can be used \cite{neudert,zali,kim04,walters,schlap,wohl,chen,jin,li,wang,qu,tang,wang2}.
Neudert \textit{et al.} have studied experimentally and theoretically the distribution of holes in the 1D cuprate
Sr$_2$CuO$_{3}$ \cite{neudert}. The authors discuss several multiband models and the effect of several terms. Recently, angle-resolved
photoemission experiments have been carried out in a related 
doped compound Ba$_{2-x}$Sr$_x$CuO$_{3+\delta}$ and
analyzed on the basis of a one-band Hubbard model with parameters 
chosen {\it ad hock} \cite{chen}.
The need to add nearest-neighbor attraction or phonons to fit 
the experiment has been suggested \cite{chen,wang,tang,wang2}.
Li \textit{et al.} studied a four-band and a one-band Hubbard
model and noted that the latter lacks the electron-hole asymmetry observed in resonant inelastic x-ray
scattering experiments \cite{li}.

The questions we want to address in this work are: 1) which is the appropriate multiband Hubbard model $H_m$ to describe
Ba$_{2-x}$Sr$_x$CuO$_{3+\delta}$? 2) What are the physical values of the parameters? 3) To what extent can this model be represented by simpler one-band ones?

Due to the large Hilbert space of $H_m$, different low-energy reduction procedures have been used to obtain simpler effective Hamiltonians for the 2D cuprates \cite{fedro,simon,brt,fei,bel,Belinicher94,ali94}. 
Most of them are based on projections of $H_m$ onto the
low-energy space of Zhang-Rice singlets (ZRS) \cite{ZR}. In spite of some
controversy remaining about the validity of this approach \cite{ebra,hamad2,adol,hamad1,jiang,ali20}, the resulting effective models seem
to describe well the physics of the 2D cuprates. 
However, the effect of excited
states above the ZRS, often neglected, can have an important role\cite{Li2021}. For
example, if one considers the Hubbard model as an approximation to $H_m$
(as done in Ref. \onlinecite{chen}), it is known that it leads, in second-order in the hopping $t$, to a term which in one dimension (1D) takes the form

\begin{equation}
H_{t^{\prime \prime }}=t^{\prime \prime }\sum\limits_{i\sigma } \left(
c_{i+2 \bar{\sigma}}^{\dagger }c_{i+1\sigma }^{\dagger } c_{i+1\bar{\sigma}%
}c_{i\sigma }-c_{i+2\sigma }^{\dagger }n_{i+1\bar{\sigma}}c_{i\sigma }+\text{%
H.c.}\right) ,  \label{hpp}
\end{equation}
with $t^{\prime \prime }=t^{2}/U>0$, where $n_{i\sigma }=c_{i\sigma
}^{\dagger }c_{i\sigma }$ and $\bar{\sigma}=-\sigma .$ This term is an
effective \emph{repulsion} and inhibits superconductivity in 1D, while as
expected, it favors superconductivity if the sign is changed \cite{ammon,lema}. Interestingly, some derivations of the generalized $t-J$ model for
2D cuprates suggest that $t^{\prime \prime }$ can be negative for some
parameters of $H_m$ \cite{ali94,sitri}, and a very small term $t^{\prime
\prime }=-t/20$ can have a dramatic effect favoring $d$-wave superconductivity \cite{bati97}. 
Even if the realistic $t^{\prime \prime }$ is positive,
it is expected to be smaller than that derived from the Hubbard model and might explain why studies of the superconductivity in the one-band Hubbard model conclude 
that part of the pairing interaction is missing\cite{dong}. 

Therefore, a discussion on the appropriate model to describe the 1D cuprates and, in particular, Ba$_{2-x}$Sr$_{x}$CuO$_{3+\delta }$ seems necessary. In this work we calculate the hopping 
parameters of the multiband model for this compound and use this information to derive simpler one-band
models.

The paper is organized as follows. In Section \ref{hm} we explain the multiband Hubbard model $H_m$ and derive its hopping parameters using maximally localized Wannier functions (MLWF).
In Section \ref{gtj} we describe the resulting 
generalized $t-J$ model obtained from  $H_m$  by a low-energy
reduction procedure explained briefly in the appendix.
In Section \ref{hone} we explain the corresponding results
for the one-band Hubbard model. In Section \ref{num} we
calculate the photoemission spectrum using the time-dependent density-matrix renormalization-group method\cite{white2004a,daley2004,Feiguin2005,Paeckel2019} and we compare it with previous experimental and theoretical results.
Section \ref{sum} contains a summary and discussion.

\section{The multiband Hubbard model}

\label{hm}

\begin{figure}
 \includegraphics[scale=0.6,keepaspectratio=true]{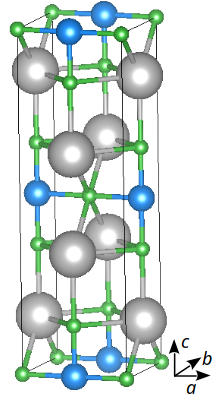}
 \caption{\label{fig1}Unit cell of BaCuO$_3$. The gray/blue/green ball are Ba/Cu/O respectively. The lattice parameters are $a=3.85$\AA, $b=4.17$\AA\  and $c=13.18$\AA. The CuO chains are along the $a$ direction with distance among them of 4.14\AA.}
\end{figure}

We use the following form of the Hamiltonian

\begin{eqnarray}
H_{m} &=&U_d \sum\limits_{i}d_{i\uparrow }^{\dagger }d_{i\uparrow
}d_{i\downarrow }^{\dagger }d_{i\downarrow }+\sum\limits_{i\sigma
}\{\epsilon _{\text{Cu}}d_{i\sigma }^{\dagger }d_{i\sigma }  \notag \\
&&+\frac{\epsilon _{\text{O}}}{2}\sum\limits_{\delta }p_{i+\delta \sigma
}^{\dagger }p_{i+\delta \sigma } 
+\epsilon _{\text{O}}^{\text{ap}}\sum\limits_{\gamma }p_{i+\gamma \sigma }^{\dagger }p_{i+ \gamma \sigma} 
\notag \\
&&+[d_{i\sigma }^{\dagger }(t_{pd}^{x}\sum\limits_{\delta }p_{i+\delta
\sigma }+t_{pd}^{y}\sum\limits_{\gamma }p_{i+\delta \gamma })  \notag \\
&&-t_{pp}\sum\limits_{\delta \gamma }p_{i+\delta \sigma }^{\dagger }p_{i+
\gamma \sigma }+\text{H.c.}]\},  \label{h3b}
\end{eqnarray}
where $d_{i\sigma }^{\dagger }$ ($p_{j\sigma }^{\dagger }$) creates a hole
with spin $\sigma $ at Cu (O) site $i$ ($j$). We choose the chain direction
as $x$ ($a$ in Fig. \ref{fig1}) and $\delta =\pm a\mathbf{\hat{x}}/2$ denote the vectors that connect
a Cu atom with their nearest O atoms in the chain direction, where $a$ is
the Cu-Cu distance that we take as $1$ in what follows. 
$\gamma $ has a
similar meaning for the apical O atoms, displaced from the chain in the $y$
direction ($c$ in Fig. \ref{fig1}). The relevant O orbitals are those pointing towards their nearest
Cu atoms. To simplify the form of the Hamiltonian, we have changed the signs
of  half of the orbitals in such a way that sign of the hopping terms do
not depend on direction and $t_{pd}^{x},t_{pd}^{y},t_{pp}>0$ \footnote{See for example red orbitals in Fig. S1 of the supplemental
material of. Ref. \onlinecite{hamad1}}.

In comparison with previous approaches \cite{neudert,li}, two terms are missing:
the intratomic O repulsion $U_p$ and the interatomic 
Cu-O repulsion $U_{pd}$. Although the former is rather sizeable
($U_p \sim 4$ eV has been estimated 
in 2D cuprates \cite{hybe}), we find
that it has very little influence on the parameters of the 
one-band models because of the low probability of double hole occupancy at the O sites. The value of $U_{pd}$ is 
difficult to determine from spectroscopic 
measurements \cite{shes} and its effect on different
quantities can be absorbed in other parameters \cite{neudert}.
We obtain a better agreement with the measured
photoemission spectra assuming $U_{pd}=0$. We also take 
$U_d=10$ eV and 
$\Delta=\epsilon _{\text{O}}-\epsilon _{\text{Cu}}=3.5$ eV, 
from calculations in the 2D cuprates \cite{hybe} and 
$\epsilon _{\text{O}}^{\text{ap}}-\epsilon _{\text{O}}=-0.4$ eV was determined as the value that leads to a ratio 1.225 between the 
occupancy of apical and chain O atoms, very similar to 
determined experimentally in Sr$_2$CuO$_3$ \cite{neudert}.
The values of the hopping parameters
$t_{pd}^{x}=1.10$ eV, $t_{pd}^{y}=1.04$ eV, and
$t_{pp}=0.60$ eV
were determined from density functional theory (DFT) calculations  along with the MLWF method.

For the DFT calculations, we use the \texttt{QUANTUM ESPRESSO} code \cite{qe,qe2}, with the GGA approximation for the exchange and correlation potential and PAW-type pseudopotentials.
The energy cut for the plane waves is 80 Ry, and the mesh used in reciprocal space is $15\times15\times5$.
The unit cell is an orthorhombic structure with lattice parameters $a=3.85$\AA, $b=4.17$\AA\  and $c=13.18$\AA, and contains two formula units, see 
Fig. \ref{fig1}.

\begin{figure}
 \includegraphics[width=0.5\textwidth,keepaspectratio=true]{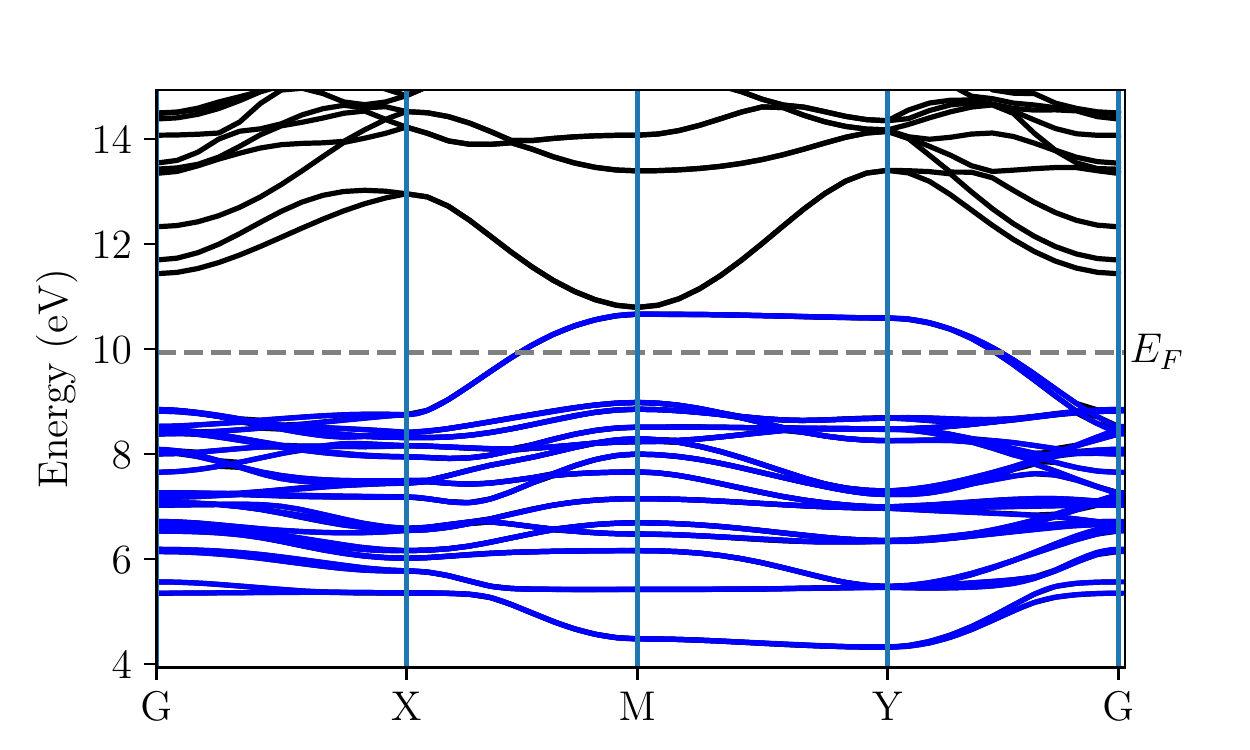}
 \caption{\label{fig2}Band structure for the BaCuO$_3$ obtained for unpolarized DFT calculation. In blue the bands from the MLWF procedure superposed with DFT ones.}
\end{figure}

We consider the spin unpolarized case and obtain the bands shown in Fig. \ref{fig2}.
The MLWF procedure involves band fitting of the DFT results, as shown in blue in Fig. \ref{fig2}.
The energy window selected to project the Wannier orbital is between 3.75 eV and 10.75 eV, and the orbitals are centered in Cu and O atoms with $d$ and $p$ character, respectively.
Other convergence parameters are also successfully evaluated, as suggested in Ref. \onlinecite{marzari}.
Finally, the hopping parameters are extracted from the Hamiltonian expressed in the Wannier basis.


\section{The generalized $t-J$ model}

\label{gtj}

Using the cell-perturbation method \cite{fei,bel}, 
with appropriate modifications for this 1D compound, we
find that the system can be described with the following generalized $t-J$ model 

\begin{eqnarray}
H_{tJ} &=&-t\sum\limits_{i\sigma }\left( c_{i\sigma }^{\dagger }c_{i+1\sigma
}+\text{H.c.}\right) \nonumber \\
&&-t_2\sum\limits_{i\sigma }\left( c_{i\sigma
}^{\dagger }c_{i+2\sigma }+\text{H.c.}\right)   \notag \\
&&+\sum\limits_{i}\left( J\mathbf{S}_{i}\cdot \mathbf{S}_{i+1} + Vn_{i}n_{i+1} \right)    + H_{t^{\prime \prime }},  \label{htj}
\end{eqnarray}
with $t_2=t/5$. Minor terms of magnitude below 0.04 eV were neglected.
The parameters of the model are given by analytical expressions in terms of
the eigenstates and energies of a cell Hamiltonian, that are obtained after
solving a $6\times 6$ matrix and two $3\times 3$ matrices. 
A summary of the method is included in the appendix. 

For the parameters of the multiband model described above, 
we obtain $t=0.443$ eV, $J=0.314$ eV, $V=-0.143$ eV, 
and $t^{\prime \prime }=0.068$ eV. Interestingly, our values
for $J$ and $t^{\prime \prime }$ without adjustable
parameters are similar to those corresponding to the Hubbard model chosen to
explain the experiments by Chen \textit{et al.} \cite{chen}. 

The larger
values of $t$ and $J$ compared to the 2D cuprates 
(for example $t=0.37$ meV, $J=0.15$ meV for T-CuO \cite{hamad1})
are expected due to the
larger overlap between the normalized O orbitals 
$\sum\limits_{\delta}p_{i+\delta \sigma }$ 
that hybridize with the Cu for nearest-neighbor Cu
positions. This leads to a larger overlap between non-orthogonal 
ZRS \cite{ali94,zhang} and to a larger extension of the orthogonal
oxygen Wannier functions centered at the Cu sites (see appendix).
For Sr$_2$CuO$_3$, the reported values of $J \sim 0.24$ meV \cite{neudert,zali,kim04,walters,schlap} are also larger than
those of 2D cuprates. 
The resulting value of $t$ is somewhat smaller than 
that used in Ref. \onlinecite{chen} but is is compensated
by the hopping to second nearest neighbors.

The fact that the nearest-neighbor attraction $V$ is larger than $J/4$ as expected for the mapping from the Hubbard to the $t-J$ model is due to the contribution of excited
local triplets absent in the Hubbard model. 

For other parameters of $H_m$, in particular increasing the ratio 
$t_{pp}/t_{pd}$ and the difference between O and Cu on-site energies, $t^{\prime \prime }$ changes sign as expected from
calculations in 2D cuprates \cite{ali94,sitri}. For example 
increasing $t_{pp}$ to 1 eV and both 
on-site energy differences to 7 eV 
(unrealistic for Ba$_{2-x}$Sr$_x$CuO$_{3+\delta}$ 
but near to the values expected for nickelates), we
obtain $t=0.531$ eV, $J=0.105$ eV, $V=-0.096$ eV,  and $t^{\prime
\prime }=-0.019$ eV, due to the increasing relative importance of excited triplets.

\section{The effective one-band Hubbard model}

\label{hone}

The generalized $t-J$ model discussed above describes the movement of
ZRS (two-hole states) in a chain of singly occupied cells. If the cells
with no holes are also considered (because for example one is interested
in larger energy scales), one can also derive a one-band Hubbard-like 
model using the cell perturbation method. A simple version of this model
has the form \cite{fedro,simon,brt}

\begin{eqnarray}
H_{H} &=&-t\sum\limits_{i\sigma }\left( c_{i\sigma }^{\dagger }c_{i+1\sigma
}+\text{H.c.}\right) [t_{AA}(1-n_{i\bar{\sigma}})(1-n_{i+1\bar{\sigma}}) 
\notag \\
&&+t_{BB}n_{i\bar{\sigma}}n_{i+1\bar{\sigma}}+t_{AB}(n_{i\bar{\sigma}}
+n_{i+1\bar{\sigma}}-2n_{i\bar{\sigma}}n_{i+1\bar{\sigma}})]  \notag \\
&&+U\sum\limits_{i}n_{i\uparrow }n_{i\downarrow }.  \label{hhub}
\end{eqnarray}

As for $H_{tJ}$, we map the ZRS into empty states of $H_H$. Then 
$t_{AA}$ coincides with $t$ of $H_{tJ}$. From the mapping procedure
we obtain 
$t_{AA}=0.443$ eV, $t_{AB}=0.421$ eV, $t_{BB}=0.369$ eV, and $U=2.083$ eV.
The model is electron-hole symmetric if and only if 
$t_{AA}=t_{BB}$,
while the photoemission of $H_m$ is asymmetric in general \cite{li}.

Another shortcoming of $H_H$ is that if the model is reduced to a generalized $t-J$ one by eliminating double occupied sites, the effective
$J=4 t_{AB}^2/U=0.376$ eV and $t^{\prime \prime}=J/4=0.094$eV  are overestimated
with respect to the values obtained in the previous section: $J=0.314$ eV,
and $t^{\prime \prime }=0.068$ eV. Instead, the NN attraction
$-J/4$ is underestimated ($V=-0.143$ eV above). This is due to the neglect of  the triplets  in $H_H$ 
(see appendix and Ref. \onlinecite{sitri}).
Therefore $H_{tJ}$ is more realistic to describe the photoemission
spectrum of hole doped 1D cuprates, unless additional terms
are added to $H_H$.

\begin{figure}
 \includegraphics[width=0.45\textwidth,keepaspectratio=true]{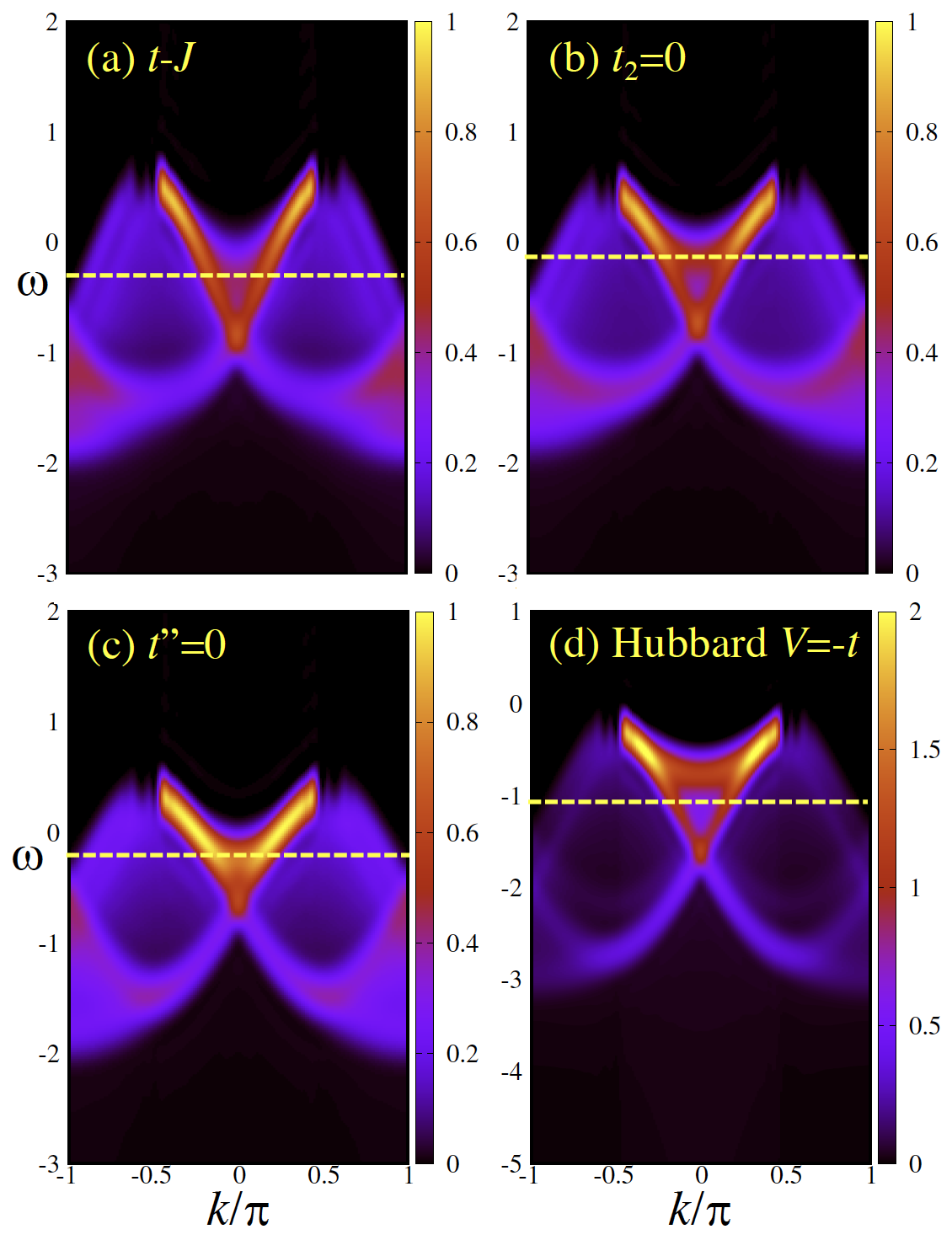}
 \caption{\label{fig3} Photoemission spectrum of (a) the generalized $t-J$, Eq.~\ref{htj}; (b) same but setting the 
 second-neighbor hopping to zero; (c) $t''=0$. Panel (d) shows results for the extended Hubbard model with $t=0.6, U/t=8; V/t=-1$ for comparison. }
\end{figure}

\begin{figure}
 \includegraphics[width=0.4\textwidth,keepaspectratio=true]{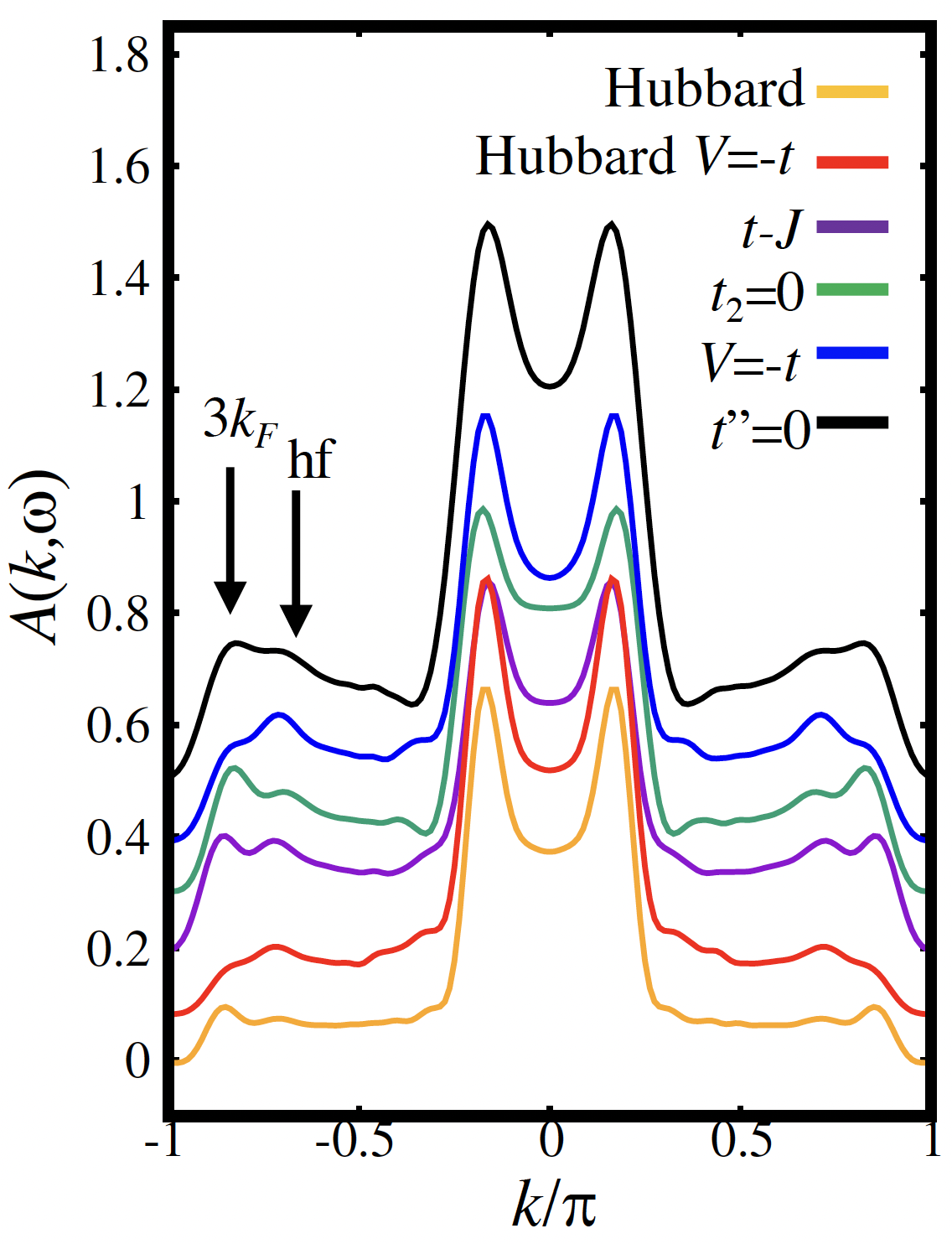}
 \caption{\label{fig4} Momentum distribution curves: cuts along the fixed frequency dashed lines in Fig.~\ref{fig3}. The features corresponding to the holon-folding (hf) and $3k_F$ bands are highlighted by arrows. The curves are ordered from bottom to top and shifted 0.1 up in intensity from the previous one for clarity.}
\end{figure}

\section{Photoemission spectrum of $H_{tJ}$}
\label{num}

\subsection{Photoemission intensity as a function of wave vector}

\label{photo}

While the effective Hamiltonian $H_{tJ}$ is enough to accurately describe the {\em energy spectrum} of $H_m$ at low energies, 
this is not the case for the spectral {\em intensity} since one needs to map
the {\em operators} for the creation of Cu and O holes in
$H_m$ to the corresponding ones of the effective low-energy Hamiltonian that one uses \cite{bati93,erol}

In the 2D cuprates, it has been found by numerical diagonalization of small
clusters, that the photoemission intensity due to O atoms at low energies
can be well approximated by the expression \cite{erol}

\begin{equation}
I_{O}=1.22\times Z(\mathbf{k})[\sin ^{2}(k_{x}/2)+\sin ^{2}(k_{y}/2)],
\label{io}
\end{equation}
where $Z(\mathbf{k})$ is the quasiparticle weight of the generalized $t-J$
model. The dependence on wave vector can be understood from the fact that
at $k=0$, the O states which point towards their nearest Cu atoms are odd
under reflection through the planes perpendicular to the orbitals, while 
the low-energy orbitals that form the ZRS are even under those reflections.
Comparison of this expression to experiment is very good \cite{hamad2}. A
variational treatment of a spin-fermion model for the cuprates also leads to
a vanishing weight at $k_{x}=k_{y}=0$ \cite{ebra}. A similar dependence is
expected in the 1D case due to the contribution of the O orbitals along
the chain, which increases the relative weight for $k_x \sim \pi/2$. To
estimate the relative weight due to these orbitals, we have
calculated the probability of creating a hole $(e^{ik}p_{i+\delta }^{\dagger
}-p_{i-\delta }^{\dagger })/\sqrt{2}$ (the minus sign is due to the choice
of phases in $H_m$) in a singly occupied cell leaving a ZRS, and also the
corresponding result for apical O and Cu. 

In addition, the observed total intensity depends on the cross sections $f$ for Cu and O, which in turn
depend sensitively on the frequency of the radiation used. 
The ratio of cross sections for the reported energy (65 eV) of the x-ray beam (available at https://vuo.elettra.eu/) is 
$f_{\text{Cu}}/f_{\text{O}}=3.077$. Using this result and 
the above mentioned probabilities for the parameters of $H_m$
described in the previous Section, we obtain that the wave vector 
dependence of the photoemission intensity can be written as

\begin{equation}
I(k)\sim Z(k)[A+B\sin ^{2}(k/2)],  \label{inten0}
\end{equation}
where $Z(k)$ is the quasiparticle weight of the generalized $t-J$ model, and 
$A=1.097$ and $B=0.205$.

\subsection{Numerical results}

We calculated the photoemission spectrum of the generalized $t-J$ model using time-dependent density-matrix renormalization group (tDMRG) \cite{white2004a,daley2004,vietri,Paeckel2019}. The simulation yields the single particle two-time correlator 
$G(x,t)=i\langle c^\dagger_\sigma(x,t) c_\sigma(L/2,0)\rangle$. This is Fourier transformed to frequency and momentum, allowing one to retrieve the spectral function as $A(k,\omega)=- \text{Im} {G(k,\omega)}/\pi$. The method has been extensively described elsewhere \cite{vietri,Paeckel2019} and we hereby mention some (standard) technical aspects. Simulations are carried out using a time-targeting scheme with a Krylov expansion of the evolution operator\cite{Feiguin2005}. Since open boundary conditions are enforced and the chain length is even, the correlations in real space are symmetrized with respect to the ``midpoint'' $x=L/2$. In order to reduce boundary effects we convolve $G(x,t)$ with a function that decays smoothly to zero at the ends of the chain and at long times, automatically introducing an artificial broadening. The function of choice is the so-called ``Hann window'' $(1+\cos{(x\pi/\sigma)})/2$, where $\sigma$ is the window width. We apply this window to the first and last quarter of the chain. We study systems of length $L=80$ sites, and $N=74$ electrons, corresponding to a $7.5\%$ doping, using $m=600$ DMRG states (guaranteeing a truncation error below $10^{-6}$), a time step $\delta t=0.05$ and a maximum time $\sigma=t_{\max}=20$. This density is chosen to maximize the holon folding (hf) effect discussed in Ref.~\onlinecite{chen}. 

Due to the one-dimensionality, the low-energy physics of the models discussed here falls into the universality class of Luttinger liquid theory\cite{GiamarchiBook,Gogolin,EsslerBook,Haldane1981}. Accordingly, excitations are not full-fledged Landau quasi-particles and the spectrum displays edge singularities instead of Lorentzians. In addition, and most remarkably, they realize the phenomenon known as spin-charge separation, with independent charge and spin excitations that propagate with different velocities and characteristic energy scales: the spinon bandwidth is determined by $J$, while the holon bandwidth by the hopping $t$. In the photoemission spectrum, these excitations appear as separate branches between $-k_F$ and $k_F$, with the spinon branch looking like an arc connecting the two points. Unlike non interacting systems, the photoemission spectrum extends beyond $|k|>k_F$ due to momentum transfer between spinons and holons: An electron with energy $\epsilon(k)$ can fractionalize into spin and charge excitations such that $\epsilon_s(q)+\epsilon_c(k-q)=\epsilon(k)$, leading to a high energy continuum and additional branches leaking out from $k=\pm k_F$ and $k=\pm 3k_F$ \cite{Ogata1990,Penc1997b,Benthien2007} (the first one is referred-to as the holon-folding band in Ref.~\onlinecite{chen} and as ``shadow bands'' in Ref.~\onlinecite{Favand1997}). 

Results for the generalized $t-J$ model [Eq.~(\ref{htj})] are shown in Fig.~\ref{fig3}(a) for the physical parameters corresponding to Ba$_{2-x}$Sr$_x$CuO$_{3+\delta}$, as discussed above. To understand the contributions of the different terms we also considered the cases without second-neighbor hopping in Fig.~\ref{fig3}(b), and without correlated hopping ($t''=0$), Fig.~\ref{fig3}(c). In all these curves, the spectral density has been rescaled according to Eq.~(\ref{inten0}). As a reference, we also show results for the extended Hubbard chain with $U=8t$ and second neighbor attraction $V=-t$ ($t=0.6eV$). 

We notice that the Hubbard chain has more spectral weight concentrated on the holon branches, while in the $t-J$ model it is more distributed in the continuum and even in the continuation of the holon bands at high energies. In addition, we observe that the $t-J$ model has a larger spinon velocity with a wider spinon branch, and a larger charge velocity with a holon band $\sim 20\%$ wider than the one for the Hubbard model (we measure the holon bandwidth as the distance between the Fermi energy and the crossing of the two holon branches at $k=0$). Since the value of $J$ remains unchanged, we attribute these effects to kinematic sources (the extra hopping terms).

In order to compare to previous attempts to interpret the experimental observations, we analyze momentum distribution curves at fixed frequency values, plotted in Fig.~\ref{fig4}, corresponding to the yellow dashed lines in Fig.~\ref{fig3}. We include results for the extended Hubbard model with $U/t=8, V=-t$ and $V=0$, that agree very well with similar previous calculations \cite{tang,wang}. 
Note that the parameters of this Hubbard model are not those that correspond to 
the mapping discussed in section IV, but the chosen value of $U$ gives rise to an effective $J$ similar to the correct one.
In this figure it is easier to observe the signatures of the $k_F$ and $3k_F$ holon branches at high momentum, which are quite faint in the color density plot and are highlighted here with arrows. One can also appreciate the qualitative differences between the extended Hubbard model with $V=-t$ and the other cases. In particular, by comparing to the standard Hubbard model with $V=0$ we notice a transfer of weight from the edge of the continuum ($3k_F$-band) to the holon-folding band, which in Ref.~\onlinecite{chen} is attributed to a phonon induced attraction. On the other hand, the $t-J$ model realizes a more prominent feature at $3k_F$ and the hf-band, and the continuum contains markedly more spectral weight in the sidebands than the Hubbard model. We also include results for the generalized $t-J$ model with $V=-t$ and we observe results practically identical to those for the extended Hubbard model with attraction.

Our results indicate that there are both kinetic as well as many-body effects that affect the relative spectral weight concentrated in the sidebands: While $t_2$ and $t^{\prime \prime }$ shift weight from the center toward the folding and $3k_F$ bands, the attraction $V$ shifts weight toward the center and from the $3k_F$ band into the 
holon-folding (hf) one. However, comparing with the cases of vanishing $t_2$ and $t^{\prime \prime }$, we see that these terms also have an effect of shifting weight from the $3k_F$ to the hf band but of smaller magnitude. 

\section{Summary and discussion}

\label{sum} 

We have started our description of CuO$_3$ chains of 
Ba$_{2-x}$Sr$_x$CuO$_{3+\delta}$ from a four-band model (with one relevant orbital per Cu or O atom). The hopping parameters of the model were obtained using maximally localized Wannier functions. Extending the cell-perturbation method used for CuO$_2$ planes of the superconducting cuprates to this one-dimensional compound, we derive simpler one-band models that are more amenable to numerical techniques due to the smaller Hilbert space. In order to account for the effect of excited triplets, the one-band Hubbard model should be supplemented by other terms not usually considered. In addition the hopping term depends on the occupancy of the sites 
involved. For energies below the value of the effective Coulomb repulsion $U$, it is more convenient to use the generalized $t-J$ model.

We have calculated the photoemission spectrum of this model using time-dependent density-matrix renormalization group. The results are in semiquantitative agreement with experiment. We obtain that the hopping to second nearest-neighbors and the three-site term $H_{t^{\prime \prime }}$ have a moderate effect in shifting weight
from the $3 k_F$ peak to the holon-folding branch, but a nearest-neighbor attraction 
has a stronger effect. For energies below $U$ and if only either electron or hole doping is of interest, a Hubbard model with an artificially enlarged $U$ that leads to the correct value of the effective nearest-neighbor exchange $J$, shows a photoemission spectrum very similar to the corresponding results for the generalized $t-J$ model.

\section*{Acknowledgments}

We thank Alberto de la Torre and Giorgio Levy for information regarding Cu and O cross sections for photoionization.  We enjoyed fruitful discussions with Alberto Nocera, Steven Johnston and Yao Wang. AAA acknowledges financial support provided by PICT 2017-2726 and PICT 2018-01546
of the ANPCyT, Argentina. AEF acknowledges support from the U.S. Department of Energy, Office of Basic Energy Sciences under grant No. DE-SC0014407. CH acknowledges financial support provided by PICT 2019-02665 of the ANPCyT, Argentina.
\appendix

\section{Derivation of the effective one-band models}

\label{deriv}

Here we summarize the application of the cell-perturbation method \cite%
{bel,fei}, to the case of the one-dimensional compound.

The Cu orbitals at each site are hybridized with symmetric linear
combinations of O orbitals of the form (dropping for the moment the spin
subscripts)

\begin{equation}
a_{i}^{\dagger }=\frac{p_{i+\gamma }^{\dagger }+p_{i-\gamma }^{\dagger }}{%
\sqrt{2}},q_{i}^{\dagger }=p_{i+\delta }^{\dagger }+p_{i-\delta }^{\dagger },
\label{qi}
\end{equation}
To change the basis of the $q_{i}^{\dagger }$ to orthonormal Wannier functions 
\cite{ZR}, we Fourier transform

\begin{equation}
q_{k}^{\dagger }=\frac{1}{\sqrt{N}}\sum\limits_{l}e^{-ikl}q_{i}^{\dagger }=%
\frac{2\cos (k\delta )}{\sqrt{N}}\sum\limits_{j}e^{-ikj}p_{j}^{\dagger },
\label{qk}
\end{equation}
where  the sum over $l(j)$ runs over all Cu(O) sites. The operators

\begin{equation}
\pi _{k}^{\dagger }=\frac{1}{2|\cos (k\delta )|}q_{k}^{\dagger },
\label{pik}
\end{equation}%
satisfy $\{\pi _{k_{1}}^{\dagger },\pi _{k_{2}}\}=\delta _{k_{1},k_{2}}$.
Transforming  to real space \ one obtains the Wannier O orbitals centered at
the Cu sites

\begin{eqnarray}
\pi _{l}^{\dagger } &=&\frac{1}{\sqrt{N}}\sum\limits_{k}e^{ikl}\pi
_{k}^{\dagger }=\sum\limits_{j}A(l-j)p_{j}^{\dagger },  \notag \\
A(j) &=&\frac{1}{N}\sum\limits_{k}e^{ikj}\text{sgn}\left[ \cos (k\delta )%
\right] =\frac{(-1)^{j-1/2}}{j\pi }.  \label{pil}
\end{eqnarray}%
Changing the basis of the O orbitals, the hopping terms in the Hamiltonian
Eq. (\ref{h3b}) (those proportional to $t_{pd}^{x},t_{pd}^{y},t_{pp}$) that
act inside each cell that includes a Cu site and the O Wannier functions
centered at the same site, becomes

\begin{eqnarray}
H_{\text{hop}}^{\text{intra}} &=&\sum\limits_{i\sigma }[d_{i\sigma
}^{\dagger }(V_{x}\pi _{i\sigma }+V_{y}a_{i\sigma })+V_{O}\pi _{i\sigma
}^{\dagger }a_{i\sigma }+\text{H.c.}],  \notag \\
V_{x} &=&2A(1/2)t_{pd}^{x},\text{ }V_{y}=\sqrt{2}t_{pd}^{y},  \notag \\
V_{O} &=&-2\sqrt{2}A(1/2)t_{pp},  \label{hopi}
\end{eqnarray}%
while the remaining part of the hopping takes the form

\begin{eqnarray}
H_{\text{hop}}^{\text{inter}} &=&\sum\limits_{i\sigma }\sum\limits_{l\neq
0}B_{l}[\pi _{i+l\sigma }^{\dagger }(t_{pd}^{x}d_{i\sigma }-\sqrt{2}%
t_{pp}a_{i\sigma })+\text{H.c.}],  \notag \\
B_{l} &=&A(l+1/2)+A(l-1/2).  \label{inter2}
\end{eqnarray}

The on-site terms of $H_{3b}$ retain the same form. This part
and $H_{\text{hop}}^{\text{intra}}$ is solved exactly in the subspaces of
one and two holes. For one hole and given spin, one has a $3\times 3$
matrix, and we denote as $E_{1}$ the lowest energy in this subspace. For two
holes and neglecting $U_{p}$ there is a $6\times 6$ matrix for the singlet
states and a $3\times 3$ matrix for each spin projection of the triplet
states .  The ground state of the subspace of singlets with energy $E_{s}$
is identified as the Zhang-Rice singlet (ZRS) \cite{ZR} and mapped into an
empty site in the effective generalized $t-J$ model.  The Coulomb repulsion
in the effective Hubbard model is $U=E_{s}-2E_{1}.$

An advantage of the cell perturbation method is that most of the hopping
terms are included in $H_{\text{hop}}^{\text{intra}}$ and including exactly
in these matrices. The rest of the hopping $H_{\text{hop}}^{\text{inter}}$
is treated in perturbation theory. The first-order correction gives rise to
effective hopping at different distances. An important difference with the
two-dimensional case is that the larger overlap between linear combinations of the original O orbitals centered at a Cu site 
[as the $q_{i}^{\dagger }$ in
Eq. (\ref{qi})] leads to larger effective hoppings and to a 
slower decay with distance.
Note that the ratio of second nearest-neighbor (NN) hopping 
to the first NN one
is $t_{2/}/t_{1}=-B_{2/}/B_{1}=1/5$ (the minus sign comes from restoring the
original signs of half of the orbitals, which have been changed to simplify $%
H_{3b}$) and for third NN  $t_{3/}/t_{1}=B_{3/}/B_{1}=3/35$ (these ratios
change if corrections due to $U_{p}$ and $U_{pd}$ are included). 

In the
effective Hubbard model, there are actually three different hopping terms
depending on the occupancy of the sites involved, while in the generalized 
$t-J$ model, only the one related with the exchange of Zhang-Rice singlets
with singly occupied sites is important.  

The most important second-order corrections in $H_{\text{hop}}^{\text{inter}}
$ lead to a superexchange $J$ and nearest-neighbor attraction $-V$  in the
generalized $t-J$ model. For example, one of these second-order processes
leads to an effective spin-flip process between a state with  one hole  with
spin $\uparrow $ at site $i$ and another with spin $\downarrow $ at site 
$i+1$, and another one with the spins interchanged, through an intermediate
state with no holes at site $i$ and two holes at site $i+1$. While the
states with one hole correspond to the ground state of the above mentioned 
$3\times 3$ matrix, the two-hole part of the intermediate states include all
singlet and triplet states of the corresponding $6\times 6$ and $3\times 3$
matrices. This is an important difference with the Hubbard model, because in
the latter, only the ground state of the $6\times 6$ matrix of singlets is
included in the effective exchange $J_{H}=4t_{AB}^{2}/U$ and the triplets
are neglected, leading to an overestimation of $J_{H}$, because the
contribution of the triplets is negative.

The next important second-order corrections in $H_{\text{hop}}^{\text{inter}}
$ correspond to three-site terms. They lead for example to an effective
mixing between states with one hole at sites $i$ and $i+1$ and a ZRS at site 
$i+2$ and states with a ZRS at site $i$ and one hole at sites  $i+1$ and  $%
i+2$. As before, performing second-order perturbation theory in the Hubbard
model includes only a few of these contributions.

The different terms can be expressed analytically in terms of the
eigenstates and eigenenergies of the matrices of the local cell mentioned
above. The expressions are lengthy and are not reproduced here.

%\bibliography{references}


\end{document}